\documentclass[letterpaper,12pt]{JHEP3}
\usepackage{bm}
\newcommand{\<}{\langle}
\renewcommand{\>}{\rangle}
\renewcommand{\d}{\partial}

\title{Magnetic permeability of near-critical 3d abelian Higgs model and
duality}

\author{Dam T.~Son\thanks{Address after March 1, 2002: Institute for
Nuclear Theory, University of Washington, Seattle WA 98195,
USA.}\\
Physics Department, Columbia University\\ 
New York, NY 10027, USA, and\\
RIKEN BNL Research Center, Brookhaven National Laboratory\\ 
New York, NY 11973, USA\\
Email: \email{son@phys.columbia.edu}
}

\abstract{The three-dimensional abelian Higgs model has been argued 
to be dual to a scalar
field theory with a global $U(1)$ symmetry.  We show that this
duality, together with the scaling and universality hypotheses,
implies a scaling law for the magnetic permeablity $\chi_m$ near the
line of second order phase transition: $\chi_m\sim t^\nu$, where $t$
is the deviation from the critical line and $\nu\approx0.67$ is a
critical exponent of the $O(2)$ universality class.  We also show that
exactly on the critical lines, the dependence of magnetic induction on
external magnetic field is quadratic, with a proportionality
coefficient depending only on the gauge coupling.  These predictions
provide a way for testing the duality conjecture on the lattice in the
Coulomb phase and at the phase transion.}

\preprint{CU-TP-1047}

\begin{document}

\section{Introduction}
\label{sec:intro}

Understanding phase transitions in gauge theories is important for the
physics of the early Universe~\cite{RubakovShaposhnikov} and heavy ion
collisions~\cite{Rajagopal}.  In some cases, e.g., electroweak theory
with a small ratio of the Higgs mass to the $W$ mass, the phase
transition can be treated reliably using perturbative techniques
\cite{ArnoldEspinosa}.  In many other cases (e.g., in the electroweak
theory with $m_H/m_W\sim1$ or in QCD) perturbative calculations are
unreliable and one has to resort to numerical simulations and other
non-perturbative methods to learn about the nature of the phase
transitions.

It is thus instructive to investigate simpler models where the phase
transitions can be studied in detail.  The abelian Higgs model (AHM,
also called the Ginzburg-Landau model), which describes the
metal-superconductor transition, is an example of a simple theory with
a rather nontrivial phase diagram.  This model has two distinct
phases: the Higgs phase, where the gauge boson (photon) is massive,
and the Coulomb phase with a massless photon.  The two phases must be
separated by a phase transition.  The temperature-induced phase
transition is first order deep in the type-I regime ($m_H\ll m_W$), as
shown by perturbative calculations~\cite{ArnoldEspinosa}, but becomes
second order as one goes to the type-II regime ($m_H\gtrsim m_W$).
The latter has been demonstrated by direct numerical
simulations~\cite{Bartholomew,KKLP,MHS}.

That the phase transition in the AHM can be second order is somewhat
surprising, given that it is always first order in $4-\epsilon$
dimensions with small $\epsilon$~\cite{HLM}.  This fact, as has been
argued, might have connection with a duality picture, according to
which the three-dimensional AHM allows a dual description as a theory
of a complex scalar field.  The role of the elementary scalar in the
dual theory is played by the Abrikosov-Nielsen-Olesen (ANO) vortex of
the AHM~\cite{BMK,DasguptaHalperin,Kleinert,KovnerRosenstein}.
Although the exact form of the dual lagrangian is not known,
quantitative predictions of the duality picture are possible near the
second order phase transition, where only the symmetries of the dual
lagrangian are important.  If the duality is valid, certain quantities
in the AHM must behave singularly near the phase transition with the
critical exponents of the $O(2)$ universality class (i.e., of the $XY$
model).  In this way one can test the duality picture on the lattice.
Numerical tests of this sort have been carried out in the Higgs phase,
where, according to duality, the tension of an ANO vortex is equal to
the mass of the dual scalar, and hence approaches zero as $t^\nu$,
where $t$ is the distance to the critical line and $\nu\approx0.67$ is
a critical exponent of the $XY$ model.\footnote{The critical behavior 
of the photon mass is discussed in ref.~\cite{HerbutTesanovic}.}
Lattice results are still
inconclusive, some appear to be inconsistent with this
prediction~\cite{KKLP,lattice}.

In this paper, we suggest some other tests of the duality hypothesis.
In addition to measuring of the vortex tension, we propose to consider
the AHM in its Coulomb phase and at the phase transition.  What one
should measure in the Coulomb phase is the magnetic permeability
$\chi_m$, which goes to zero as one approaches the critical line where
the Meissner effect (i.e., the Higgs mechanism) starts taking place
and the system is perfectly diamagnetic.  We shall show that $\chi_m$
is proportional to the square of the decay constant of the Goldstone
boson in the dual theory.  This mapping is precise (provided duality
is valid) and involves only quantities which are not renormalized.
Using scaling and universality hypotheses, we then show that the
critical behavior for $\chi_m$ is $\chi_m\sim t^{\nu}$.  Exactly on
the critical line, the magnetic permeability vanishes and the
dependence of magnetic induction $B$ on external magnetic field $H$ is
nonlinear.  We shall demonstrate, by using simple scaling arguments,
that this dependence is quadratic: $B\sim H^2$, with a proportionality
coefficient depending only on the gauge coupling $e$.

The paper is organized as follows.  In section~\ref{sec:review} we
review the duality picture.  Section~\ref{sec:arg} is devoted to the
study of the magnetic permeability of the Coulomb phase.  The main
line of logic in this section consists of three steps.  In the first
step (section~\ref{sec:chimchi}) one relates the magnetic permeability
of the AHM with the susceptibility of the dual vacuum to the $U(1)$
chemical potential.  The second step (section~\ref{sec:chif})
establishes, in the dual theory, the connection between the
susceptibility with the decay constant of the Goldstone boson.  The
third step (section~\ref{sec:ft}) determines the critical behavior of
the decay constant.  Each step involves fairly well-known arguments,
but we believe their synthesis is new.  In section~\ref{sec:BH2} we
discuss the response of the AHM to an external magnetic field exactly
on the critical line.  Section~\ref{sec:concl} contains concluding
remarks.

\section{Review of the duality picture}
\label{sec:review}

Although we are mostly interested in the phase transition driven by
temperature, thanks to dimensional reduction we can describe the
static long-distance physics by an Euclidean three-dimensional AHM
theory.  Changing the temperature in the (3+1)d theory corresponds
varying the parameter of the 3d dimensionally reduced theory.  We
shall thus start directly from the AHM in three spatial dimensions.
It is sometimes useful, especially in our discussion of duality, to
turn one spatial dimension into a temporal dimension; we then have a
(2+1)d AHM, where the ANO vortex is a particle.  This particle is the
elementary scalar in the dual theory.  The duality was first argued by
using a formal representation of the partition function of the AHM in
terms of loops~\cite{BMK,DasguptaHalperin,Kleinert}.  Subsequently it
has been given an operator form in 2+1
dimensions~\cite{KovnerRosenstein}.

\TABULAR{|c|c|}{\hline
abelian Higgs model            & complex scalar theory\\
\hline
magnetic induction (total magnetic field) & $U(1)$ charge density\\
external magnetic field        & chemical potential\\
\hline
Coulomb phase                  & broken $U(1)$ (superfluid phase)\\
photon                         & Goldstone boson\\
Higgs particle                 & global vortex\\
magnetic permeability $\chi_m$ & square of decay constant $f^2$\\
\hline
Higgs phase                    & unbroken $U(1)$ (Mott insulator)\\
Abrikosov-Nielsen-Olesen vortex & scalar particle\\
vortex tension                 & scalar mass\\
critical magnetic field        & scalar mass\\
\hline}
{The duality maps between the abelian Higgs model and the dual theory
of a complex scalar.  Some of the mappings are explained further in
the paper.\label{table:duality}}

We summarize the correspondence between the AHM and the dual complex
scalar theory in table~\ref{table:duality}.  The Higgs phase is dual
to the phase with unbroken $U(1)$ symmetry (the Mott insulator phase),
and the Coulomb phase is dual to the phase where the $U(1)$ global
symmetry is broken (the superfluid phase) by the condensation of the
dual scalar field (``vortex condensation'').  The dual of the massless
photon in the AHM Coulomb phase is the superfluid Goldstone boson.
This is possible because photon has only one polarization in 2+1
dimensions.  The vortices in the AHM and the scalar theory, when
exist, correspond to a particle in the other theory.

For the purpose of this paper, the most important equation of the
duality picture comes from the identification of the magnetic field in
the AHM with the $U(1)$ current in the dual theory,
\begin{equation}
   {e\over2\pi} \epsilon^{\mu\nu\lambda} F_{\nu\lambda} = j^\mu .
  \label{duality}
\end{equation}
The conservation of the $U(1)$ current is an automatic consequence of
eq.~(\ref{duality}).  The correspondence (\ref{duality}) implies that
the magnetic flux is proportional to the $U(1)$ charge:
$eF_{12}/2\pi=j_0$, where the proportionality coefficient $e/2\pi$ is
such that the quantum of the magnetic flux corresponds to an unit
charge.  This suggests that the vortex of the AHM is the elementary
scalar particle in the dual theory.

If one works in 2+1 dimensions, it is also possible to construct in
the AHM a local operator creating unit magnetic flux, which plays the
role of the order parameter in the dual
theory~\cite{KovnerRosenstein}.  The precise form of this local order
parameter is not as important to us as the fact that it exists.  We
denote the order parameter as $V$; the Mott insulator phase has
$\<V\>=0$ and the superfluid phase has $\<V\>\neq0$.  The exact form
of the lagrangian for $V$, ${\cal L}(V)$, is not know and also not
important for further discussion.  On the other hand, the expression
for the $U(1)$ charge density operator is known exactly,
\begin{equation}
  j_0 = -i(\pi V-\pi^*V^*)\,,
\end{equation}
where $\pi$ is the operator canonically conjugate to $V$.

\section{Magnetic permeability of the Coulomb phase}
\label{sec:arg}

\subsection{Magnetic permeability and susceptibility of the dual vacuum}
\label{sec:chimchi}

We now turn on an external magnetic field $H$ in the AHM.  Assuming
that the external field is aligned along the $z$ axis, the lagrangian
becomes
\begin{equation}
  {\cal L} = {\cal L}_0 - H F_{12}\,.
\end{equation}
From the point of view of the physics in (2+1)d, the hamiltonian is
changed by
\begin{equation}
  {\cal H} = {\cal H}_0 - H F_{12}\,.
\end{equation}
Once $H$ is turned on, $B\equiv F_{12}$ obtains an expectation value.
Following established tradition, we shall call the total magnetic
field $\<B\>$ the magnetic induction, and $H$ the external magnetic
field.  The magnetic permeability is defined as $\chi_m=\d\<B\>/\d
H|_{H=0}$.

In the (2+1)d dual theory, $B$ is proportional to the charge density.
Thus, the external magnetic field in the AHM corresponds to the
chemical potential coupled to the $U(1)$ charge in the dual
theory~\cite{FisherLee},
\begin{equation}
  {\cal H}(\pi,V) = {\cal H}_0(\pi, V) + i\mu (\pi V - \pi^* V^*)\,,
  \label{Hmu}
\end{equation}
where ${\cal H}_0(\pi,V)$ is the hamiltonian in the absence of the
chemical potential.  The relation between the chemical potential in
the dual theory and the external magnetic field in AHM, according to
eq.~(\ref{duality}), is
\begin{equation}
  \mu = {2\pi\over e} H\,.
  \label{muH}
\end{equation}
The magnetic permeability of the AHM is thus proportional to the
susceptibility of the vacuum of the dual theory with respect to the
$U(1)$ chemical potential,
\begin{equation}
  \chi_m \equiv {\d\<B\>\over\d H} = 
  {4\pi^2\over e^2}{\d \<j_0\> \over\d\mu}
  \equiv {4\pi^2\over e^2}\chi\,.
  \label{chim}
\end{equation}
Therefore, in order to find the critical behavior of $\chi_m$, one
needs to find that of $\chi$ in the dual theory.  Notice that
$e^2\chi_m$ receives no renormalization in the AHM, and $\chi$ is not
renormalized in the dual theory.

Before turning to the Coulomb phase, which is the main subject of this
paper, let us make a comment about the response of the Higgs phase to
an external magnetic field.  According to the duality hypothesis, the
Higgs phase of the AHM corresponds to the Mott insulator phase of the
dual theory where the $U(1)$ symmetry is not spontaneously broken.
This phase has a mass gap $m$ equal to the mass of the elementary
scalar.  At zero temperature, a chemical potential less than $m$ has
no effect on the vacuum, since creating an excitation costs positive
energy ($m-\mu$ for a particle and $m+\mu$ for an antiparticle).  In
particular, the charge density remains zero if $|\mu|<m$.  The
counterpart of this is the Meissner effect in the AHM: the total
magnetic field inside a superconductor remains zero if one turns on a
small external magnetic field.  If the chemical potential in the dual
theory exceeds $m$, then it becomes energetically favorable to create
particles.  This corresponds, in the AHM, to the existence of a
critical magnetic field (more precisely, the lower critical magnetic
field $H_{c1}$), above which the magnetic field begins to penetrate
the type-II superconductor.  

\subsection{Susceptibility and decay constant of the Goldstone boson}
\label{sec:chif}

Let us now return to the Coulomb phase, or the superfluid phase from
the dual point of view.  Due to the breaking of the $U(1)$ symmetry,
the low-energy dynamics of the theory contains only one Goldstone
mode, which is the phase $\varphi$ of the order parameter
$V=|V|e^{i\varphi}$.  Let us first assume no external magnetic field
(chemical potential in the dual theory).  In contrast to the
lagrangian for $V$ which is not known, the form of the low-energy
effective lagrangian for the Goldstone mode is completely fixed,
\begin{equation}
  {\cal L}_{\rm eff}(\varphi) = {f^2\over2}(\d_\mu\varphi)^2\,,
  \label{Leff}
\end{equation}
where $f$ is some parameter to be called the ``decay constant'' of the
Goldstone boson, since it is the analog of the pion decay constant
$f_\pi$ in the chiral lagrangian of QCD.\footnote{In condensed matter
literature $f^2$ is called the stiffness.}  Notice, however, that in
$d$ dimensions $f^2$ has dimension $d-1$, i.e, in 3d $f^2$ has the
dimension of energy.  The effective theory~(\ref{Leff}) is valid below
the typical energy of non-Goldstone excitations, which we denote as
$m_\sigma$.

If the chemical potential $\mu$ is small compared to $m_\sigma$, its
effect can be captured within the framework of the effective theory.
Moreover, the way $\mu$ enters the effective lagrangian can also be
fully determined~\cite{Stephanov}.  First we notice that, by taking
the Legendre transform of eq.~(\ref{Hmu}), the lagrangian for $V$ in
the presence of a chemical potential can be obtained from the
lagrangian with no chemical potential ${\cal L}_0(V)$ by making the
replacement
\begin{equation}
  \d_0 V \to \d_0 V - i\mu V\,.
\end{equation}
In other words, $\mu$ enters the lagrangian as the zeroth component of
a fictitious gauge field arising from gauging the global $U(1)$
symmetry.

A useful trick is to consider this fictitious gauge field as a fully
dynamical field $A_\mu$ and make the substitution $A_0=\mu$, $A_i=0$
at the very end~\cite{Stephanov}.  The lagrangian for $V$ is invariant
under gauge transformations $V\to e^{i\alpha}V$, $A_\mu\to
A_\mu+\d_\mu\alpha$.  If $A_\mu$ is small and slowly varying, this
gauge invariance must also be a property of the effective lagrangian
for $\varphi$ as well.  Noticing that $\varphi$ transforms as
$\varphi\to\varphi+\alpha$, one sees that the gauge invariance can be
preserved if in the effective lagrangian~(\ref{Leff}) $\d_\mu\varphi$
is replaced by $\d_\mu\varphi-A_\mu$.  Substituting in the final
answer $A_\mu=(\mu,{\bf 0})$, one discovers the effective lagrangian
for the Goldstone mode in the presence of the chemical potential:
\begin{equation}
  {\cal L}_{\rm eff}(\varphi) = 
  {f^2\over2}[(\d_0\varphi-\mu)^2-(\d_i\varphi)^2]\,.
  \label{Leffmu}
\end{equation}
The ground state free energy (i.e., pressure) is obtained by putting
$\varphi=\mbox{const}$ into eq.~(\ref{Leffmu}), and is equal to
$f^2\mu^2/2$.  The susceptibility is, by definition, the second
derivative of the pressure with respect to the chemical potential
$\mu$, i.e.,
\begin{equation}
  \chi = f^2.
  \label{chif}
\end{equation}
Together with eq.~(\ref{chim}), the magnetic permeability of the AHM
is now related to the decay constant of the Goldstone boson.

\subsection{Critical behavior of the decay constant}
\label{sec:ft}

Now let us find out the critical behavior of $f^2$.  This can be done
by using various arguments.  The simplest one is purely dimensional in
nature: one notices that $\varphi$ is a dimensionless phase variable,
defined mod $2\pi$, so it is not renormalized.  Therefore $f^2$ has
the canonical dimension $d-2=1$.  According to the scaling hypothesis,
the mass scale of non-Goldstone excitation $m_\sigma$ is the only
dimensionful scale near the critical point, so $f^2\sim
m_\sigma^{d-2}$, i.e., $f^2\sim t^{(d-2)\nu}=t^\nu$ in 3d.

Another argument is similar to the one originally used by
Josephson~\cite{Josephson} for the stiffness parameter of superfluid
helium, and in ref.~\cite{critical_pions} for the pion decay constant
near the chiral phase transition.  Let us decompose the complex order
parameter $V$ into the real and imaginary parts,
\begin{equation}
  V = V_1 + i V_2\,,
\end{equation}
and choose the ground state so that $\<V_1\>=\<V\>$ and $\<V_2\>=0$.
At distances large compared to $m_\sigma^{-1}$, the amplitude of the
order parameter is effectively frozen, and $V_2$ is proportional to
the Goldstone field, $V_2=\<V\>\varphi$.  Therefore, the correlator of
$V_2$ can be found from the effective lagrangian for $\varphi$,
eq.~(\ref{Leff}),
\begin{equation}
  \int\!d^3x\, e^{-iq\cdot x} \< V_2(x)V_2(0) \> = 
  {\<V\>^2\over f^2q^2}\,,
  \qquad q\ll m_\sigma\,.
  \label{smallq}
\end{equation}
On the other hand, at distances small compared to $m_\sigma^{-1}$, the
system is effectively at the critical line and is $O(2)$ symmetric.
In this regime the correlator of $V_2$ have the same form as that of
the order parameter at the Wilson-Fisher fixed point, i.e.,
\begin{equation}
  \int\!d^3x\, e^{-iq\cdot x} \< V_2(x)V_2(0)\> =
  \int\!d^3x\, e^{-iq\cdot x} \< V_1(x)V_1(0)\> = {c\over q^{2-\eta}}\,,
  \qquad q\gg m_\sigma\,,
  \label{largeq}
\end{equation}
where $c$ is a constant independent of the distance $t$ to the
critical line.  The two formulas (\ref{smallq}) and (\ref{largeq}) are
valid in two opposite regimes, but must smoothly match to each other
at $q\sim m_\sigma$.  From this condition one finds the critical
behavior of $f^2$,
\begin{equation}
  f^2 \sim \<V\>^2 m_\sigma^{-\eta} \sim t^{2\beta-\eta\nu} .
\end{equation}
By using a well-known (hyperscaling) relation between the critical
exponents,
\begin{equation}
  2\beta = \nu(d-2+\eta) \,,
  \label{hyperscaling}
\end{equation}
one finally obtains $f^2\sim t^{(d-2)\nu}=t^\nu$, which agrees with
the previous dimensional argument.  Notice that $f\sim t^{\nu/2}$
scales differently from the order parameter, $\<V\>\sim t^\beta$, in
contrast to what one may naively expect, but the difference is
numerically small because of the smallness of $\eta$ in
eq.~(\ref{hyperscaling}).  Since, as we have shown in
eqs.~(\ref{chim}) and (\ref{chif}), the magnetic permeability of the
AHM is proportional to $f^2$, one concludes that $\chi_m$ approaches 0
as
\begin{equation}
  \chi_m \sim t^\nu ,
\end{equation}
which is one of the main results of this paper.

\section{Magnetic response at the critical line}
\label{sec:BH2}

Let us now turn our attention to the response of the critical AHM to
an external magnetic field.  On the critical line $\chi_m=0$, so the
dependence of the magnetic induction $B$ on the external field $H$
must be nonlinear.  The dependence of $B$ on $H$ can be found in the
most intuitive way by going to the dual theory, where it governs the
dependence of the charge density $j_0$ on the chemical potential $\mu$
at the critical line.  Since at the critical point the dual theory is
a conformal field theory with no intrinsic scale, the only
dimensionful parameter is $\mu$.  Due to charge conservation, $j_0$
has the canonical dimension, i.e., $d-1$.  Therefore,
$j_0\sim\mu^{d-1}$, which in three dimensions reads
\begin{equation}
  j_0 = C \mu^2\,,
  \label{largemu}
\end{equation}
From eqs.~(\ref{duality}), (\ref{muH}) and (\ref{largemu}), we find
that the magnetic induction is quadratic over the external magnetic
field:
\begin{equation}
  B = \left( \frac{2\pi}e \right)^3 C H^2 \,.
  \label{BH2}
\end{equation}

The dependence (\ref{largemu}) can be found by another (related)
argument, which elucidates the magnetic response of the near-critical
AHM.  Let us slightly go away from the critical line toward the
Coulomb phase.  The dual theory is in the superfluid phase,
characterized by a small decay constant $f$ and a small mass scale of
non-Goldstone excitations $m_\sigma$.  If the chemical potential $\mu$
is small compared to $m_\sigma$, then $j_0$ is linear of $\mu$, with
the proportionality coefficient equal to the susceptibility
$\chi=f^2$,
\begin{equation}
  j_0 = f^2 \mu, \qquad \mu\ll m_\sigma\,.
  \label{smallmu}
\end{equation}
In the opposite regime $\mu\gg m_\sigma$, the slight deviation from
the critical line is unimportant, and the dependence must be of the
form $j_0=C\mu^n$, where $C$ is independent of $t$ and $n$ needs to be
found.  Recalling that $f^2\sim m_\sigma^{d-2}$, this power-law
behavior can match with eq.~(\ref{smallmu}) at $\mu\sim m_\sigma$ if
and only if $n=d-1=2$, i.e., it must be of the form (\ref{largemu}).

As the infrared fixed point of the dual theory is unique, one should
expect the constant $C$ in eq.~(\ref{largemu}) to be universal.  In
this case the proportionality coefficient between $B$ and $H^2$ in
eq.~(\ref{BH2}) depends only on the gauge coupling $e$, but not on the
Higgs self-coupling (provided the phase transition is second order).

\section{Conclusion}
\label{sec:concl}

One should note that there exists a rather trivial way to test duality
in the Coulomb phase (in fact, also in the Higgs phase), which is
based on the measurement of the specific heat.  However, this method
is potentially difficult from the numerical point of view due to the
smallness of the critical exponent $\alpha$ in the $O(2)$ universality
class.  Therefore, one should look for a quantity with a stronger
critical behavior.  We propose here to use the magnetic permeability,
which goes to zero as $\chi_m\sim t^\nu$, for this purpose.  We have
demonstrated that the scaling law for $\chi_m$ can be found by
invoking fairly standard arguments of duality, scaling, and
universality.  In addition, we have shown that on the line of second
order phase transition the dependence of magnetic induction $B$ on
external magnetic field $H$ is quadratic, $B\sim e^{-3} H^2$, and the
proportionality coefficient does not depend on the Higgs
self-coupling.

These predictions, in principle verifiable on the lattice, are rather
nontrivial tests of duality.  These tests, relevant for the Coulomb
phase and the critical line, should be considered as supplements to
the measurement of the vortex tension in the Higgs phase.  Looking
from a broader perspective, we hope that the investigation of the
phase transition in the AHM will further elucidate the physics at the
phase transitions of more realistic theories, e.g., the standard model
or QCD.

\acknowledgments

I am indebted to Mikko Laine, Kimyeong Lee and Misha Stephanov for
stimulating discussions, and Hagen Kleinert, Flavio Nogueira and 
Zlatko Te\v sanovi\' c for comments on the manuscript.
I thank RIKEN, Brookhaven National
Laboratory, and U.S.\ Department of Energy [DE-AC02-98CH10886] for
providing the facilities essential for the completion of this work.
This work is supported, in part, by a DOE OJI grant and by the Alfred
P.\ Sloan Foundation.

\end{document}